\documentclass[letter]{aa}

\def\kms{${\rm km}\,{\rm s}^{-1}$}
\newcommand{\degree}{\ensuremath{^\circ}}
\def\losv{{\it v}$_{\sc {\rm los}}$}

\usepackage{graphicx,epsfig,fancyhdr,rotating,amsmath,natbib}
\begin{document}
\title{SUMER observations of the inverse Evershed effect in the transition 
region above a sunspot}
\author{L. Teriaca, W. Curdt, S.K. Solanki}

\offprints{L. Teriaca,\\ \email{teriaca@mps.mpg.de}}

\institute{Max-Planck-Institut f\"{u}r Sonnensystemforschung (MPS), 
	37191 Katlenburg-Lindau, Germany}

\date{Received xxx, 2008; accepted xxx, 2008}

\abstract
{}
{We analyse SUMER spectral scans of a large sunspot within 
active region NOAA 10923, obtained on 14 -- 15 November 2006, to 
determine the morphology and 
dynamics of the sunspot atmosphere at different heights/temperatures.}
{The data analysed here consist of spectroheliograms in the continuum 
around 142.0~nm
and in the Si~{\sc iv}~140.2~nm, 
O~{\sc iii}~70.3~nm,
N~{\sc iv}~76.5~nm,
and O~{\sc iv}~79.0~nm
spectral lines. 
Gaussian-fitting of the observed profiles provides line-of-sight velocity 
and Doppler-width maps.}
{The data show an asymmetric downflow pattern compatible with the 
presence of the inverse Evershed flow in a region within 
roughly twice the penumbral radius at
transition-region temperatures up to 0.18~MK. The motions, highly inhomogeneous 
on small scales, seem to occur in a collar of radially directed filamentary 
structures, with an average width less than the 1~Mm spatial resolution of 
SUMER and characterised by different plasma speeds.
Assuming that the flows are directed along the field lines, we deduce that
such field lines are inclined by 10\degree\, to 25\degree\, with respect to 
the solar surface.
}
{}

\authorrunning{Teriaca et al.}

\titlerunning{SUMER observations of the inverse Evershed effect}

\maketitle
%

\section{Introduction}
One of the most peculiar spectroscopic signatures of sunspots is the 
observed opposite shift and/or asymmetry of spectral lines observed in
(disc) centreward and limbward penumbral areas. The effect was
discovered 
in photospheric lines by \citet{Evershed:1909a} 
and named after him. 
In the photosphere, the line shifts 
are, ever since Evershed, generally interpreted in terms of nearly horizontal 
flows directed outward from the umbra. 
Such flows reach speeds of a few kilometres per second and appear to be mainly 
associated with dark penumbral channels when observed at high ($<1''$)
spatial resolution \citep[e.g.,][]{Wiehr-Degenhardt:1994,solanki:2003}.
In fact, the magnetic field topology is inhomogeneous on small scales, and the
exact geometry of single flow channels is still under debate.

\citet{Evershed:1909b} also gave the first hint that 
the deduced direction of motion reverses its sign in the chromosphere.
This ``inverse Evershed effect''
appears more extended than its photospheric 
counterpart, with radial velocities peaking beyond the external penumbral rim
and with a larger vertical component \citep[e.g.,][]{dere-etal:1990}. 

The persistence of the ``inverse Evershed effect'' at transition-region (TR)
temperatures was first observed by \citet{nicolas-moe:1981}. Such flows 
have greater radial speeds, exceeding by almost one order of magnitude 
those observed in the chromosphere \citep[e.g.,][]{dere-etal:1990} and reach 
their maximum further out from the penumbra. 
In fact, due to the different scales and densities involved, and due to 
the oppositely directed flow, the ``inverse Evershed effect'' has little 
direct relationship with the photospheric Evershed effect. Here we
use the term ``inverse Evershed effect'' to indicate a system of inwardly
directed flows distributed all around the spot and distinct from, but not 
necessarily unrelated to, the strong downflows associated with sunspot 
plumes.

Observations with high spatial and 
spectral resolution in TR lines also revealed many multi-component profiles 
indicating supersonic downflows in sub-arc-second fine structures above umbrae
\citep{kjeldseth-moe-etal:1993}. It is not clear whether such 
supersonic downflows are related mainly to sunspot plumes or (also) to the 
inverse Evershed flow. 

In this paper, we present a series of raster scans of a large sunspot taken over a
period of two days. The high spectral and spatial resolution of the data 
allowed us to determine the characteristics of the inverse Evershed flow in the
sunspot TR. The photospheric Evershed flow of this sunspot has been studied in 
detail by \citet{ichimoto-etal:2007}.
\section{Observations and data reduction}
Raster scans were repeatedly taken by SUMER
\citep{wilhelm-etal:1995} aboard SOHO
in two spectral regions on 14 and 15 Nov. 2006, following AR 10923 
from the time it crossed the central meridian to a longitude of about 
15$^\circ$ West. Data 
were acquired with detector~B and the 1\arcsec\,$\times$\,300\arcsec~slit.
In the short wavelength scans (type N~{\sc iv} in Table~\ref{tab1}), 
we analysed selected 0.22~nm wide spectral windows centred on the
N\,{\sc iv}~76.5~nm ($T_{\rm e}=0.15$~MK)
and O\,{\sc iv}~79.0~nm ($T_{\rm e}=0.18$~MK)
TR lines. 

A step of 1.5$''$ was commanded after each 30~s exposure.
The long wavelength (type Si~{\sc iv}) scans (exposure time 45~s and 1.12$''$ 
step size) contained 
Si\,{\sc iv}~140.3~nm ($T_{\rm e}=0.08$~MK), 
O\,{\sc iii}~70.4~nm (observed in the second order of diffraction at 
140.8~nm, $T_{\rm e}=0.11$~MK), 
and a continuum band
at 142.0~nm, which lacks any prominent emission line, both in the
first and the second order \citep{curdt-etal:2001}.
Calculations by \cite{VAL:81} indicate that
the continuum around 142.0~nm is formed in the low chromosphere at a height
of roughly 200~km above the temperature minimum between the solar photosphere
and chromosphere.
More observational details are given in Table~\ref{tab1}.
Data were reduced by applying standard procedures from SolarSoft. 
%
\begin{table}[!b]
\caption{Data summary with centre and size of scans, and with the $\theta$
 angle between the sunspot centre and the disc centre.}
\label{tab1}
\centering
\begin{tabular}{llrcrc}
\hline
\# & type        & centre/$''$ & size/$''$ &$\theta$/\degree &(Day) time/UTC\\
\hline
1a &Si\,{\sc iv} & 90,-122 &225$\times$300& 8.9  &(14) 19:24-22:05\\
1b &N\,{\sc iv}  &117,-122 &300$\times$300& 9.9  &(14) 22:26-00:56\\
2a &Si\,{\sc iv} &149,-122 &225$\times$300& 11.5 &(15) 01:27-04:07\\
2b &N\,{\sc iv}  &185,-122 &300$\times$300& 12.4 &(15) 04:07-06:37\\
3a &Si\,{\sc iv} &212,-122 &225$\times$300& 14.0 &(15) 06:57-09:37\\
3b &N\,{\sc iv}  &244,-122 &300$\times$300& 15.2 &(15) 10:02-12:32\\
\hline
\end{tabular}
\end{table}
%
   \begin{figure}[!th]
   \centering
    \resizebox{\hsize}{!}{\includegraphics{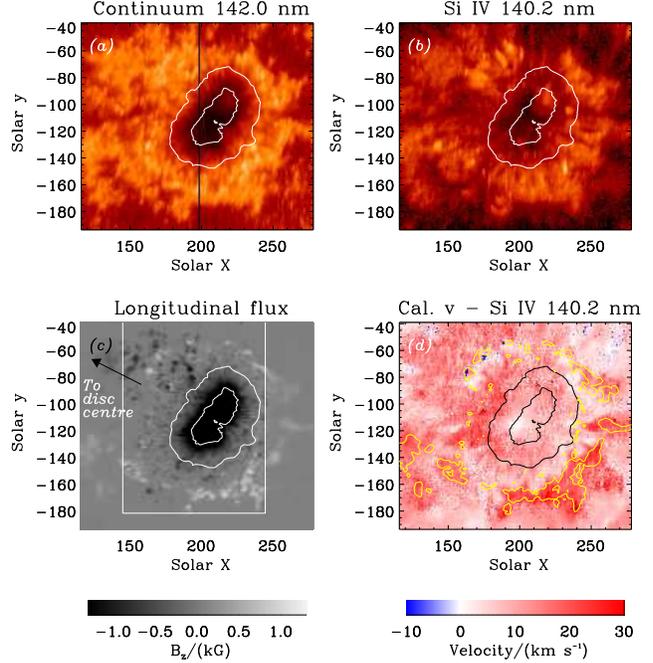}}
   \caption{Logarithmic radiance (panels {\bf a} and {\bf b}) and 
   	{\sc los} velocity (\losv) (panel {\bf d}) map obtained 
	on 15 Nov. (dataset \#3a). Panel {\bf c}
	shows the longitudinal magnetic flux from a combined Hinode/SOT (from
	11:00 UTC, area within the white box) and a MDI full-disc magnetogram 
	(from 11:15 UTC).
	Iso-contours (white and black) indicate the umbral and penumbral edges 
	as obtained from the combined SOT and MDI continuum images, while the 
	yellow contours on the Si~{\sc iv} \losv\, map outline the 
	100~G ($=10$~mT) level of 
	the positive magnetic polarity.
	Coordinates (X=0$''$, Y=0$''$) 
	correspond to solar disc centre.}
   \label{fig:ref}
   \end{figure}

Magnetograms and filtergrams of the
observed region were obtained by MDI \citep{scherrer-etal:1995} aboard SOHO
and SOT-SP \citep{tsuneta-etal:2008} aboard Hinode.
The alignment between SUMER and MDI data was performed in two steps. 
First, the solar rotation was subtracted from the raster increment.
Then, the SUMER scans in the continuum at 142.0~nm were correlated to the
nearest (in time) map of the absolute magnetic flux. We estimate the final 
alignment to be accurate within 1$''$. Finally, the SOT-SP data were aligned 
with MDI to within 1$''$.
\section{Data analysis}
The aim of the observations discussed here was to obtain a high signal-to-noise 
ratio (S/N) in the bright structures (i.e., plages, sunspot plumes) associated 
with a sunspot. However, due to the strong contrast between these bright 
structures and the quiet areas south of the sunspot, the spectral profiles in 
these darker areas have a low S/N that prevents a satisfactory analysis. 
For this reason, the data sets were selectively re-binned. 
For all profiles having less than five counts at line peak (about 25 counts in 
the line), a box of $7\times7$ profiles centred on the selected
location was defined, and the profile was replaced by the sum of all
profiles in the box having less than five counts at line top. 
The spectra were then fitted with a single Gaussian plus a linear 
background to retrieve line positions and widths.
Note that radiance maps, in contrast, are at full resolution and are obtained by
simple subtraction of the nearby continuum.
In the case of the long wavelength scans, a 300~s full detector readout
was taken directly before the scan and was used to calibrate our
wavelength scale through several lines of neutral
and singly ionised species that characterise the solar spectrum above 90~nm
\citep[see, e.g.,][]{teriaca-etal:1999a}. However, this reference spectrum 
was taken over the sunspot where chromospheric flows of 
few kilometres per second may affect this calibration. 
Hence, we estimate our zero velocity to be accurate
within $\approx\,3$~\kms.
In the short-wavelength spectra, no calibration lines are available and 
the zero velocity of a line refers to its position averaged over 
the whole raster. We also note that the
lower spectral resolving power at these wavelengths leads to larger 
uncertainties in the retrieved line parameters.
Maps of the line width ($w=c_{0}~\Delta\lambda_{1/\rm e}/\lambda$, where 
$c_{0}$ is the speed of light in vacuum) were obtained (but not shown) after 
correcting the measured line widths for the SUMER instrumental 
profile\footnote{function CON\_WIDTH\_FUNCT\_4.pro from SolarSoft.}. 
\section{Results and conclusions}
Figure~\ref{fig:ref} shows a selection of the radiance and line-of-sight 
({\sc los}) velocity (\losv) maps (dataset \#3a). 
   \begin{figure}[!t]
    \resizebox{\hsize}{!}{\includegraphics{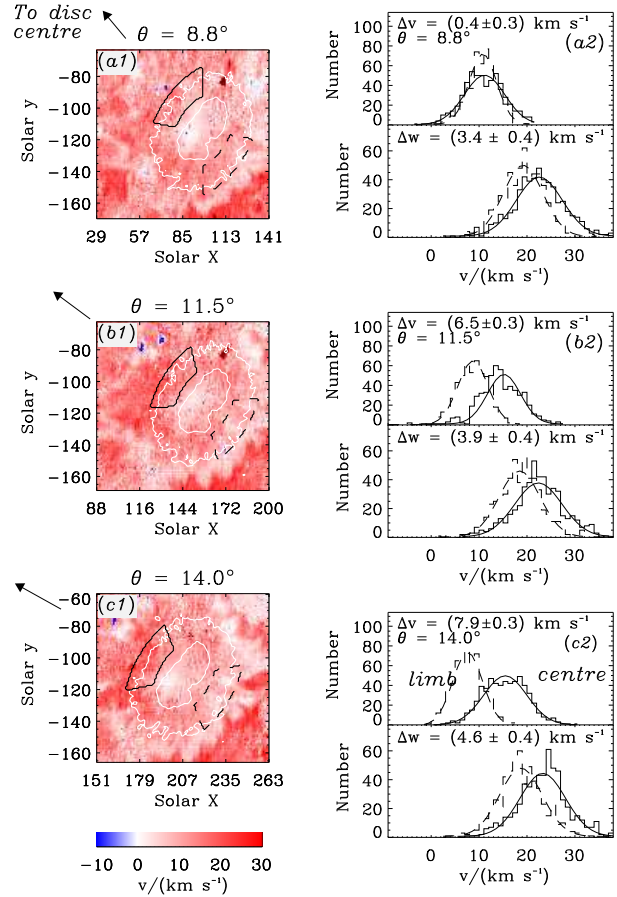}} 
   \caption{Panels {\bf a1} to {\bf c1}: extracts of the Si~{\sc iv}
        \losv\, maps from data sets \#1a, \#2a, and \#3a. White contours 
	from MDI continuum images mark inner and outer penumbral rims.
	In each panel, two 
	sectors at opposite sides along the external parts  
	of the penumbra, both 
	with an angular extension of $80\degree$, are outlined by 
	solid and dashed black lines. 
	Panels {\bf a2} to {\bf c2}
	show the velocity (upper) and width (bottom)
	distributions (histograms) within the selected regions
	together with single-Gaussian fits. The separations between the
	centroids of the centreward (cw) and limbward (lw) \losv, $\Delta v$, 
	and line widths, $\Delta w$, distributions are also given.}
   \label{fig:si4}
   \end{figure}

Closer inspection of the Si~{\sc iv} \losv\, map in 
Fig.~\ref{fig:ref} reveals the limbward (lw) edge of the penumbral 
region to be less red-shifted than the centreward (cw) edge, a 
signature compatible with the inverse Evershed effect. The first 
question to consider is why we see redshifts on both the cw and lw sides 
of the penumbra and superpenumbra, unlike the photospheric Evershed flow 
\citep[redshift on the lw side and blueshift on the cw side; e.g.,][]{solanki-etal:2006}.
One important factor is that the field lines traced by the TR emission are significantly
inclined with respect to the solar surface. Because our observations were carried out close to
disc centre ($\theta\le15$\degree), an inclination of the field lines $ > 15$\degree\ 
is sufficient to show only redshift at all positions.
The figure also shows that the 
inverse Evershed flow seems to end at patches of opposite magnetic polarity
surrounding the sunspot and particularly visible on the lw side of the sunspot.
The opposite polarity is actually related to
strong downflows that are difficult to explain in the context of the
inverse Evershed effect. (For inward flows along radially directed 
loops, upflows would be expected at the outer footpoints.) Although a detailed
investigation requires a field extrapolation and will be part of
an upcoming paper, we can already speculate that the ``inverse Evershed flows'' 
are most likely occurring along long-reaching loops that have the outer 
footpoints outside the penumbra, possibly outside the FOV. 
The emission from these long loops would dominate above the umbra and the
penumbra leading to the observed inverse Evershed signature.
Further out, plage emission would dominate over that of the long-reaching 
downflowing loops and would be 
characterised by the mostly redshifted emission typical of the solar 
TR, both quiet and active. In an
active region \citet{teriaca-etal:1999a} found an average downflow of about 
16~\kms\ in Si~{\sc iv} and O~{\sc iv}, while a value of 17.6~\kms\ was found by
\citet{achour-etal:1995} in O~{\sc iv}. These values compare well with the
$\approx$~21~\kms\ downflow (practically constant through all data sets) 
observed in the plage area south east of the spot.
The masking effect could also be enhanced by the fact that only the lower 
part of such loops could be cold enough to emit significantly in TR lines.

\begin{figure}[!t]
   \resizebox{\hsize}{!}{\includegraphics{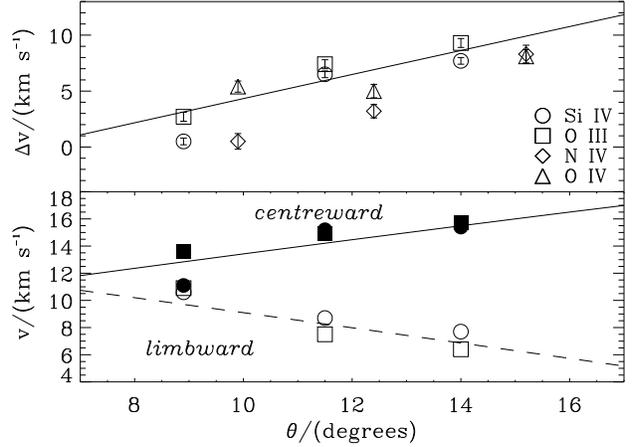}}
   \caption{{\it Top}: Differences between the positions of the centreward 
   (cw) and the
   limbward (lw) \losv\, distributions as a function of the $\theta$ angle.
   {\it Bottom}: Positions of the cw (full) and lw (open symbols) \losv\,
   distributions of Si~{\sc iv} and O~{\sc iii}. The lines are
   obtained assuming a flow velocity of 33~\kms\, along field lines with
   an average inclination (on the solar surface) of 14\degree\, in the cw
   direction and 26\degree\, in the lw direction.}
   \label{fig:all}
   \end{figure}

Extracts from Si~{\sc iv} \losv\,
maps around the spot for datasets \#1a, \#2a, and \#3a are shown in
Fig.~\ref{fig:si4} confirming these findings and showing that the effect appears 
to increase with increasing $\theta$ angle, as expected.
To study this aspect, two areas along the external part of the penumbra
were selected in each \losv\, and line width map. 
As an example, for each of the Si~{\sc iv} \losv\,
maps in Fig.~\ref{fig:si4}, the two selected regions are
outlined on panels 
from {\it a1} to {\it c1}. Distributions of the Si~{\sc iv} \losv\,
(top of panels {\it a2} to {\it c2}) and line widths (bottom of
panels {\it a2} to {\it c2}) within these areas were computed for each
dataset. The separation between cw and lw
distributions increases with increasing $\theta$ angle. 
Figure~\ref{fig:all} shows that the difference between the central positions 
of Gaussian fits to the cw and the lw \losv\, distributions for all
lines
increase, to first order linearly, as a function of $\theta$.

Important is that the lw line width distributions are shifted towards smaller 
values than the cw distributions (Fig.~\ref{fig:si4}). 
If $\Delta(w)$ is the distance 
between the centroids of the width distributions then, at $\theta=14$\degree,
$\Delta w=(4.6\pm0.4)$~\kms\, for Si~{\sc iv} and 
$\Delta w=(4.0\pm0.4)$~\kms\, for O~{\sc iii}.
Moreover, we note in Fig.~\ref{fig:si4} that
the \losv\, distributions are narrower on the lw side. 
However, the average standard 
uncertainties of \losv\, in the cw distributions are greater than those in the 
lw distributions (e.g., 3~\kms\, versus 2.1~\kms\, for Si~{\sc iv})
because of lower average radiances in the cw sector. This leads to greater
broadening of the cw distribution.
The ``de-noised'' widths of the \losv\, cw ($\delta v_{\rm cw}$)
and lw ($\delta v_{\rm lw}$) distributions, obtained by 
subtracting the squared uncertainty from the squared measured distribution 
widths, are roughly comparable: 
$\delta v_{\rm cw} - \delta v_{\rm lw}=(0.4\pm1)$~\kms\ for Si~{\sc iv} 
and $\delta v_{\rm cw} - \delta v_{\rm lw}=(1.8\pm0.6)$~\kms\ for O~{\sc iii} 
(average and standard deviation of the differences at the three
angular distances).

We explain these results by considering that the cw field 
lines form a smaller angle with the {\sc los} direction than the lw field
lines. We now assume that inverse Evershed motions occur along multiple
thin flux tubes roughly aligned with 
each other \citep[in analogy to the picture of the photospheric penumbral 
field,][]{solanki-montavon:1993,schlichenmaier-etal:1998}. If each of these is
characterised by a different velocity, then the integral over several 
tubes (also along the {\sc los} for an optically thin plasma)
would naturally produce a broader distribution of velocities when looking at
an angle closer to the main direction of the field lines. 
If the flux-tube diameters are smaller than the 1~Mm of the SUMER spatial
resolution, then the associated sub-resolution velocity distributions 
contribute to the observed line widths that become larger when looking more 
parallel to the field lines.
If the flow structures are resolved by SUMER, we expect 
$\delta v_{\rm cw} - \delta v_{\rm lw} \gg \Delta w \approx 0$,
while we expect $\Delta w \gg \delta v_{\rm cw} - \delta v_{\rm lw} \approx 0$
for unresolved structures.
Our results, given above, are consistent with the 
second case, indicating that the inverse Evershed flow is inhomogeneous at 
scales below the SUMER spatial resolution of 1~Mm.

The knowledge of the absolute velocities (\losv ) of Si~{\sc iv} and 
O~{\sc iii} allows estimating the average field inclination
(see Fig.~\ref{fig:all}) that, assuming a flow velocity of 33~\kms,
results to be about 14\degree\, in the cw direction and 26\degree\, in the 
lw direction. The relatively large inclination at the lw side also explains
why we see only reduced redshift on the lw side of the penumbra and not blueshift. 
The latter would eventually become visible when the spot rotates more than 
26\degree\ away from disc centre.

These results confirm the inverse Evershed effect to be visible in
the TR up to temperatures of about 0.18~MK
(where O~{\sc iv} is formed). Its spatial extent appears to be no more
than twice the penumbra for the studied sunspot and is nearly the same in 
all lower and mid TR lines. The flux channels have diameters on average somewhat 
smaller than 1~Mm and an inclination relative to the solar surface that can be 
estimated as about 10\degree\, to 25\degree.

\begin{acknowledgements}
      The SUMER project is financially supported by DLR, CNES, NASA, and the
      ESA PRODEX programme (Swiss contribution).
      Hinode is a Japanese mission developed and launched by ISAS/JAXA, with 
      NAOJ as domestic partner and NASA and STFC (UK) as international 
      partners. It is operated by these agencies in co-operation with ESA and 
      NSC (Norway).
      The authors thank H. Peter and K. Wilhelm for valuable comments and 
      suggestions, and U. Sch\"{u}hle and D. Germerott for their help in 
      acquiring the data.
\end{acknowledgements}

\bibliographystyle{aa}

\bibliography{0209}

\end{document}